\documentclass[aps,pra,twocolumn,superscriptaddress]{revtex4}

\usepackage{graphicx}
\usepackage{url}
\usepackage[colorlinks=true,linkcolor=blue,citecolor=blue,urlcolor=blue]{hyperref}

\usepackage{amsmath}

\renewcommand{\section}[1]{{\bf\itshape #1}}

\begin{document}

\title{Encoding Matters: A Qubit-Efficient Reformulation for Quantum Resilient Routing}
\author{Emanuele G. Dalla Torre}
\email{emanuele.dalla-torre@quantymize.com}
\affiliation{QuantyMize Quantum Advance Ltd, Tel Aviv 6789145, Israel}
\author{Bat-El Friedman}
\affiliation{QuantyMize Quantum Advance Ltd, Tel Aviv 6789145, Israel}
\author{Daniel Porat}
\affiliation{QuantyMize Quantum Advance Ltd, Tel Aviv 6789145, Israel}
\date{\today}

\begin{abstract}
Qubits are powerful computational resources because each additional qubit doubles the size of the accessible state space. This expressive power comes at
a cost: adding qubits also increases circuit size, constraint overhead, and
exposure to hardware noise. These limitations are especially critical for
near-term quantum computers, making the number of qubits required to formulate a
problem a central practical concern. In this work, we present a method
for efficiently mapping logistics and routing problems onto qubits through structure-aware
encoding reduction. We demonstrate the method on a benchmark problem of
latency-resilient routing under dual-link failures. For this benchmark, the
reduced formulation substantially lowers both the qubit count and the compiled
circuit complexity while preserving the original optimization task. By running the reduced model on the same trapped-ion quantum computer as the benchmark, we obtain a $100\times$ improvement in the success
probability. This finding shows how careful problem
reformulation can make current quantum hardware substantially more effective for
practical optimization.
\end{abstract}

\maketitle

\section{Introduction}
Quantum optimization is motivated by the promise of using quantum superposition to explore large
combinatorial spaces faster than classical search. To translate this promise into actual quantum advantage, one needs to use the available quantum resources, such as qubit count, circuit depth, and shot count, as efficiently as possible. In particular, the number of required qubits determines the
accessible state space, yet also increases circuit
width, entangling-gate count, constraint overhead, and sensitivity to noise. For near-term hardware,
the representation of a problem can therefore matter as much as the
optimization algorithm itself
\cite{lucas2014ising,preskill2018quantum,tamura2021performance,glos2022space}.

This resource tradeoff becomes especially important when solving real-life optimization models relevant to the industrial sector. Problems in logistics, routing, scheduling, finance, and
resource allocation are usually defined by structured objectives and rich
logical relations \cite{li2020survey,volpe2025review,he2021overview,nuesslein2024less}.
Translating such problems to models that can be solved on quantum computers, such as Ising or quadratic unconstrained binary optimization (QUBO) models, is not a neutral step: the
encoding determines how many qubits are required, how many constraints must be
enforced, and how difficult the final circuit is to execute. Much of the
literature emphasizes hardware progress, error mitigation, and variational
strategies, but canonical encodings often retain variables that are fixed by
the problem, implied by local topology, or redundant once exclusion constraints
are enforced. Encoding choices can therefore have a strong downstream effect on
tractability, resource efficiency, and solution quality
\cite{lucas2014ising,tamura2021performance,glos2022space,zaman2021pyqubo,yarkoni2022quantum}.

\begin{figure}[t]
\centering
\includegraphics[width=0.95\columnwidth,trim=0.5cm 6cm 0.5cm 0.5cm,clip]{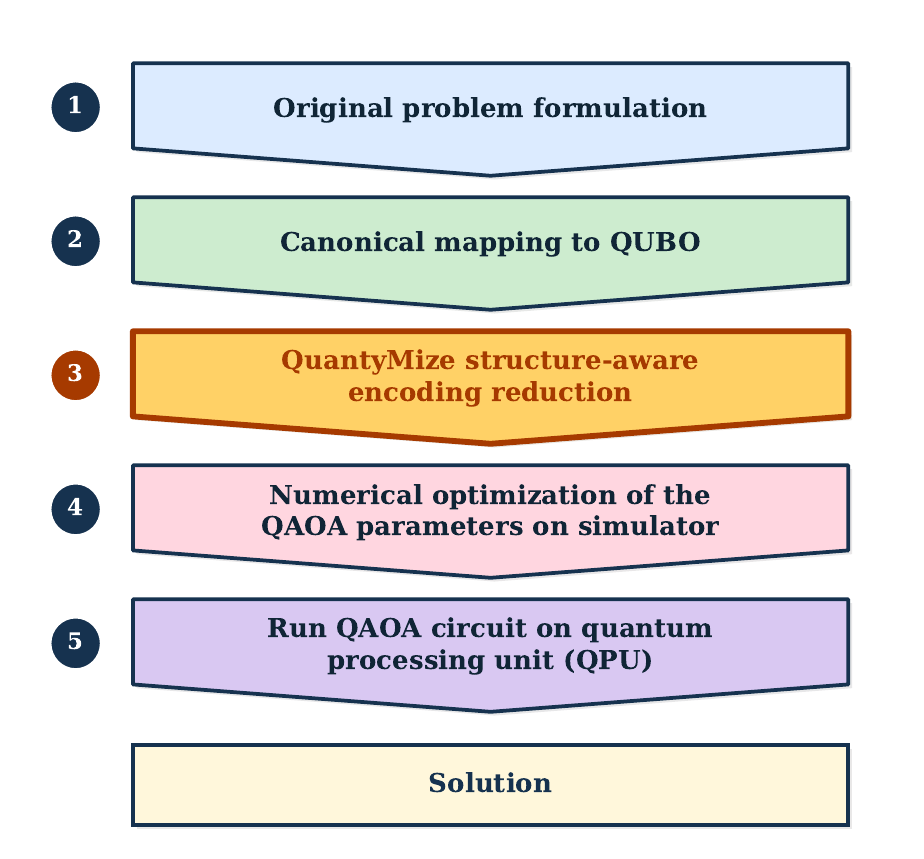}
\caption{Workflow for the proposed approach to solving network optimization problems. The highlighted step (3) represents the rule-based encoding reduction that transforms the 24-qubit canonical formulation into a 7-qubit representation, before simulation and hardware execution.}
\label{fig:workflow}
\end{figure}

In this manuscript, we introduce an efficient method for solving graph
problems on quantum computers through structure-aware encoding reduction.
Starting from a canonical node-edge formulation, the method applies
problem-driven logical rules to eliminate fixed variables, merge equivalent
variables, and simplify branching constraints before the final QUBO is built. The overall workflow is summarized in Figure~\ref{fig:workflow}. Importantly, the reduction rules are local:
they depend only on the neighborhood of each node and edge of the network, and do not require an enumeration of complete paths.

As a concrete case study, we consider the latency-resilient Layer 3 routing
benchmark introduced by Harb et al.\ \cite{harb2026quantum}. This problem is related to shortest disjoint-path
algorithms, minimum-shared-edge formulations, shortest-path variants with
disjoint counterparts, and quantum annealing approaches to traffic and network
optimization
\cite{suurballe1984quick,li1992finding,xu2006complexity,omran2013shared,krauss2020shortest,neukart2017traffic,inoue2021traffic}.
 Harb et al. formulated the optimization problem as an integer linear
program together with an equivalent QUBO model, which they solved by QAOA
\cite{farhi2014quantum,wang2018qaoa}. Their work demonstrated feasibility on
both simulator and hardware, but at a significant resource cost. For a model
network with 5 nodes, the canonical formulation requires 24 qubits and achieves a hardware
success probability of only $0.2\%$. Our approach reduces the encoding to 7 qubits and sharply lowers circuit complexity, improving both
simulator and hardware performance. 

\section{Methods}
We consider graph optimization problems in which one seeks
paths, subgraphs, or flows on an underlying graph $G=(V,E)$ subject to logical
constraints. In a canonical node-edge encoding, binary variables are introduced
to indicate whether a node or edge participates in a candidate solution. For
routing-type problems with multiple paths, this construction is repeated for
each path  (see Refs.
\cite{feld2018hybrid,irie2019quantum,seker2020routing,he2021gps,cattelan2022modeling}
for related implementations).
% Such encodings are general and transparent, but often redundant: they generically include
%variables that are fixed by the problem statement, implied by graph topology, or enforced by constraints. 
The canonical encoding is attractive because it follows directly from the graph
structure and supports a clean QUBO construction. However, it also contains
logical redundancy. Some node variables are fixed by the routing task itself,
some variables are ruled out by the requirement that the two paths remain
separate, and some node-edge relationships can be inferred locally from the
topology. 

%Our reformulation preserves the benchmark objective and path-validity
%requirements while reducing the number of independent binary variables.

Our method addresses this redundancy to improve the QUBO formulation of the model. The key idea is to start from a canonical node-edge formulation and then apply a
small number of deterministic logical rules that simplify the representation
without changing the optimization target. Motivated by Harb et al., we consider a multi-terminal network in which a source node is connected to each terminal through non-overlapping paths. We then use source and terminal conditions, exclusion constraints, and
local graph structure to derive five rules that reduce the number of logical variables. Importantly, all five rules are {\it local} and  can be applied by scanning each node and its connected edges only, so that one does not require prior knowledge of the full set of
admissible paths. 

The rules are given below in an arbitrary order and are applied iteratively to the canonical encoding until no
additional variables can be fixed, merged, derived, or excluded. While the precise definition of each rule depends on the problem at hand, our approach can be easily generalized to other network and logistic problems on a graph.

\paragraph*{Rule 1: Source and terminal nodes.}
For each path, the source node and its designated terminal node must be active.
Accordingly, the corresponding node variables are fixed to 1. This removes
degrees of freedom that are present in the canonical encoding but are not
genuine optimization choices.

\paragraph*{Rule 2: Edges attached to fixed active nodes.}
For any source or designated terminal node fixed to 1 by Rule 1, exactly one incident candidate edge must be active.
\emph{(a) Two-edge case.} If two incident candidate edges are present, one edge may be represented by a binary variable $q_1$ and the other by its
complement, $1-q_1$. \emph{(b) Three-edge case.} If instead three incident candidate edges are present, two may be represented by
independent variables $q_1$ and $q_2$, while the third is derived as
$1-q_1-q_2$. The same construction extends to higher-degree nodes.

\paragraph*{Rule 3: Complementary terminals.}
To prevent a route from passing through another route's terminal, that terminal
node is fixed to 0, together with all edge variables that directly connect
to that terminal within the current path encoding. 
%In other words, when encoding
%the path to $T_1$, the variables associated with entering $T_2$ are removed,
%and vice versa. 
%This rule hard-codes a separation condition that is otherwise
%left to the optimizer and penalty terms.

\paragraph*{Rule 4: Degree-2 internal nodes.}
If an internal node has exactly two incident candidate edges that have not been fixed to zero, the node variable and the two adjacent edge
variables are identical. Activating the node therefore activates both admissible
incident edges, while activating either edge implies that the node belongs to
the path. The three variables can be represented by a single binary degree of
freedom. This rule is especially effective after Rule 2 has pruned invalid
branches from the graph.

\paragraph*{Rule 5: Internal-node elimination by edge balance.}
For every feasible assignment, path continuity at an internal node $v$ requires $x_v=\frac{1}{2}\sum_{e\in\delta(v)}x_e$,
where $\delta(v)$ is the set of edges incident to $v$. In words, the number of activated incident edges equals two if the node is active and zero if it is inactive. Now, if substitutions from the
preceding rules reduce the right-hand side of the equation exactly to one surviving binary
variable $q$ or its complement $1-q$, the node variable is not independent and
is replaced by that binary quantity. Note that if the right-hand side corresponds to a more general expression, it is not substituted for $x_v$, because it may take values outside $\{0,1\}$. In that
case, the node variable and its consistency constraint are retained. 

\section{Benchmark implementation} 
\begin{figure}[t]
\centering
\includegraphics[width=0.82\columnwidth]{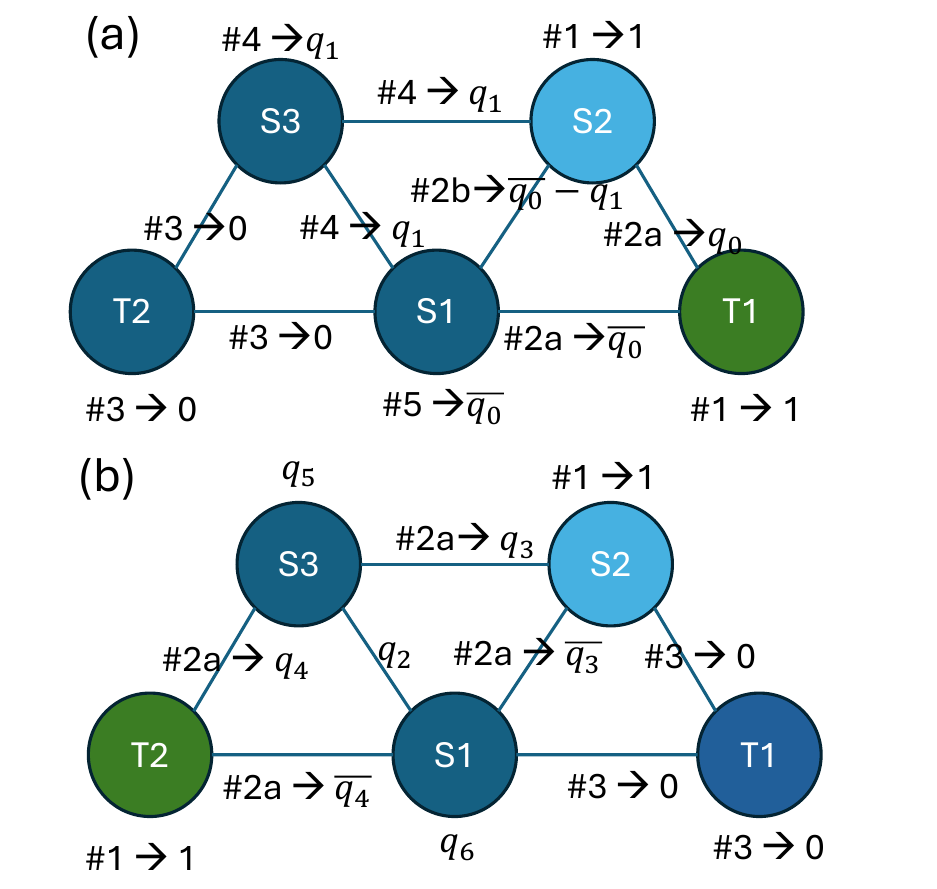}
\caption{Rule-based reduction of the canonical encoding for the 
$S_2$ benchmark case executed on quantum hardware, with paths to (a) $T_1$ and
(b) $T_2$. Each
annotation of the form ``\#N $\rightarrow q$'' denotes the application of
the $N$th reduction rule, mapping a canonical variable either to a fixed
value or to one of the surviving logical variables. The final reduced encoding
contains the seven logical variables $q_0,\ldots,q_6$ and their complementaries $\bar{q_i}=1-q_i$.}
\label{fig:mapping}
\end{figure}

We instantiated the five rules on the latency-resilient Layer 3 routing
benchmark introduced by Harb et al.\ \cite{harb2026quantum} and reproduced in Figure~\ref{fig:mapping}.  Each edge carries a delay cost, and pairs of edges may
also carry a joint-failure penalty that captures correlated failures. The
optimization objective therefore combines path length with resilience to
correlated link failure. 
%Harb et al. introduced a canonical
%node-edge formulation in which binary decision variables are introduced for both
%node membership and edge membership on each path. 
In the canonical representation used by Harb et al., for each source vertex $s \in \{S_1,~S_2,~S_3\}$, one set of variables
describes the path from $s$ to $T_1$ and a second set describes the path from
$s$ to $T_2$. The constrained routing problem is then mapped to a QUBO
representation whose cost Hamiltonian is minimized with QAOA. The
topology used in the benchmark has 5 nodes and 7 edges, giving $12$ binary variables for each
path, and a total of 24 variables.

%Variables linked by
%identity relations are represented by a single binary variable, and mutually
%exclusive branches are encoded using the reduced admissible set. The resulting
%formulation preserves the benchmark objective, namely the joint minimization of
%latency and dual-link-failure exposure, but acts on a smaller set of independent
%binary variables. The reduced binary model is then mapped to a QUBO and to the
%corresponding cost Hamiltonian for quantum optimization.

For concreteness, we consider source vertex $s=S_2$, which admits five valid
route pairs and is the correlated-failure case executed on quantum hardware in
the benchmark study. 
%By applying the above-mentioned 5 rules, we obtain a
%Since each update uses
%only local incidence information, the same reduction procedure can be deployed
%on larger graphs as a preprocessing pass before QUBO construction. 
The application of the five aforementioned rules leads to the variable-elimination map shown in the annotated graph in
Figure~\ref{fig:mapping}. In that figure, each annotation of the form ``\#N
$\rightarrow q$'' indicates that Rule $N$ maps the corresponding node or edge
variable either to a fixed value or to one of the surviving logical variables.
An overbar denotes binary complementation, $\bar q_i=1-q_i$.
The two subfigures represent the two path encodings after pruning and
substitution. When this reduction is carried through on the network at
hand, the canonical 24-variable benchmark encoding collapses to seven independent
logical variables, labeled $q_0,\ldots,q_6$. 
%In this convention, $q_4$
%denotes the $S_1$--$S_3$ edge in path (b), while $q_5$ and $q_6$ denote the
%$S_3$ and $S_1$ node activations in path (b), respectively. The remaining
%source-edge variable in path (a) is not independent, but is derived locally as
%$1-q_0-q_1$.
The resulting quantum model therefore requires 7 qubits and has a search space
of $2^7=128$ binary assignments, rather than 24 qubits and
$2^{24}\approx 1.7\times 10^7$ assignments in the canonical encoding.
For completeness, we verified the equivalence of the two formulations
by exhaustively enumerating all $2^7=128$ assignments of the reduced model.
After decoding, the valid assignments reproduce the complete set of valid routes
in the canonical formulation with unchanged objective values, including the
same optimal route and ground-state energy.

After the reduction, the remaining path-validity conditions are implemented as
local quadratic penalties. 
%The source node on path (a) has three admissible outgoing
%edges, two represented by $q_0$ and $q_1$ and the third derived as $1-q_0-q_1$; equivalently, this edge may be labeled $\bar q_0-q_1$ in the graph annotation.
%The three edges incident to $S_1$ in panel (a) then sum to
%$2(1-q_0)$, so Rule 5 identifies $x_{S_1}$ with the complemented binary
%variable $\bar q_0$ without introducing an additional node variable.
For path (a), we need to guarantee that there is exactly one outgoing edge from $S_2$ by avoiding the simultaneous activation of $q_0$ and $q_1$,
which is penalized by
\begin{align}
    C_{\mathrm{source}} = \alpha q_0 q_1 .
\end{align}
For path (b), we require the number of activated edges incident on each of the vertices $S_3$ and $S_1$ to equal twice its corresponding node-activation variable, $q_5$ and $q_6$, respectively. This constraint is implemented by the penalty function
%the activation of the internal nodes must remain explicit in order
%to allow reconnection of branches later in the graph. We therefore introduce
%the node variables $q_5$ and $q_6$, together with the internal edge variable $q_4$, and impose
\begin{align}
    C_{\mathrm{node}} = \alpha\left[(q_2+q_3+q_4-2q_5)^2\right.\nonumber\\
    \left. + (1-q_2+1-q_3+q_4-2q_6)^2\right].
\end{align}
Finally, internal vertex-disjointness between the two routed paths is enforced by multiplying the assignment variables for $S_3$ and $S_1$ on both paths:
\begin{align}
    C_{\mathrm{overlap}} = \alpha\left[ q_1q_5 + (1-q_0)q_6 \right].
\end{align}
The same coefficient $\alpha$ multiplies all four validity penalties and is chosen
large enough ($\alpha=10$) to guarantee separation between valid and constraint-violating
assignments.
These terms are added to the main cost function, which also includes the same latency and failure-penalty terms as in the benchmark.

\section{QAOA solution} We next evaluated the reduced formulation against the canonical benchmark on the same topology and failure scenarios using the same workflow as in Harb et al.: After generating the reduced QUBO, we constructed the corresponding QAOA circuit with $p$ layers and optimized its parameters on a noiseless simulator to maximize the statevector overlap with the ground-state subspace of the reduced cost Hamiltonian. To achieve this goal, we performed a multi-start search and refined the best parameter set using the bounded L-BFGS-B optimizer. 
%We defined the ideal success probability as the ground-state overlap and the hardware success probability as the fraction of measured bit strings that decode to an optimal valid route. 

%This reduction is the central structural result of the encoding stage. Relative
%to the benchmark formulation, it decreases the quantum register from 24 qubits
%to 7 qubits before any circuit-level optimization is performed. 

%After implementing the reduction, we construct the corresponding QUBO and cost Hamiltonian and optimize the QAOA parameters on the simulator. %
%The reduced encoding achieves higher ideal and hardware success probabilities with QAOA $p=4$, compared with $p=20$ for the benchmark simulation and $p=8$ for its hardware implementation. 

The improvement is already present in ideal simulation. Under the
canonical benchmark encoding, the ideal success probability with $p=20$ QAOA layers is reported as
$1.5\%$. Under our reduced encoding, the ideal success probability rises more than 10 times to
$25.4\%$, already for $p=4$.  This simulator result is important because it shows that the advantage of the
reformulation is not limited to noise suppression alone. Even before hardware
noise is taken into account, the reduced encoding changes the optimization
landscape in a favorable way. In other words, the gain is already present at the
level of representation and not only at the level of hardware execution.

The optimized $p=4$ QAOA circuit was then executed in three independent 100-shot runs on the same quantum hardware used in the benchmark, namely the trapped-ion IonQ Forte Enterprise 1 device accessed through Amazon Braket, without error mitigation. Our reduced encoding recovered the optimal solution in 60 out of 300
shots, leading to a success probability of $20\%$. 
This corresponds to a $100\times$ improvement relative to
the benchmark hardware result ($0.2\%$). This finding is consistent with the significant reduction in the number of entangling gates: our results were obtained using just 53 ZZ rotations, compared with approximately 1000 two-qubit gates in the benchmark implementation.

\section{Discussion and Conclusion}
\begin{figure}[t]
\centering
\includegraphics[width=0.5\textwidth]{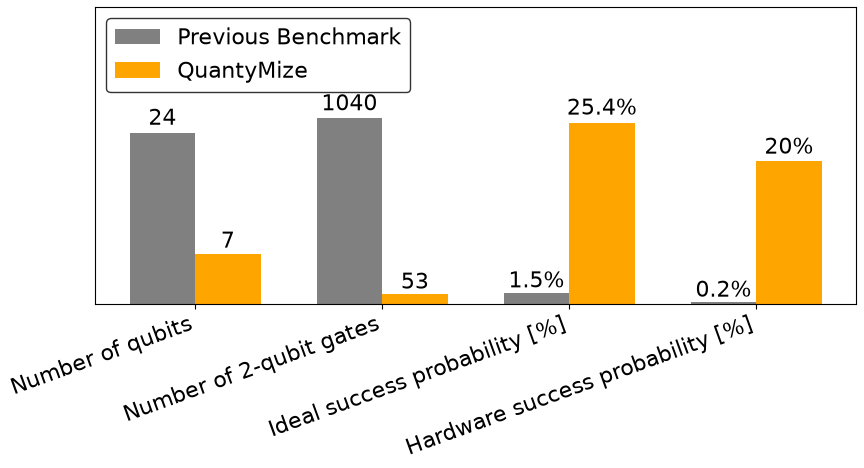}
\caption{Comparison between the canonical benchmark encoding (gray) and the proposed reduced encoding (orange). The four groups report the number of qubits, the number of two-qubit entangling gates, the ideal ground-state success probability, and the hardware optimal-solution probability, respectively.}
\label{fig:comparison}
\end{figure}

The central finding of this study is that problem formulation can be a dominant
factor in near-term quantum performance. Starting from the canonical benchmark
encoding, we introduced a rule-based reduction that preserves the routing
objective while substantially reducing the required quantum resources. As summarized in Figure~\ref{fig:comparison}, our reduced formulation improves all key indicators with respect to the benchmark and, in particular, increases the hardware success
probability by two orders of magnitude.

% showing that careful reformulation alone
%can materially improve near-term quantum performance.

These results indicate that the main limitation of near-term applications of quantum computers
is not simply hardware noise in isolation. Rather, the encoding itself
substantially determines how much of that noise the algorithm must tolerate.
Reducing the number of logical variables lowers the size of the Hilbert space to
be explored, while reducing the number of entangling gates directly lowers the
opportunity for error accumulation. The improvement therefore reflects both a
better optimization representation and a better match between the problem model
and the capabilities of current gate-based quantum devices.

For telecommunications applications, this result suggests that useful quantum
network optimization may become practical earlier through better encodings than
through hardware progress alone. In the resilient-routing setting considered
here, the reformulation transforms a weak and rare hardware signal into a
substantially more recoverable one. This is important for operational settings
in which solution reliability, shot efficiency, and compute cost all matter.

The broader implication extends beyond routing. Many near-term quantum
optimization problems in logistics, energy, and finance are currently limited by
the cost of representing classical constraints on noisy hardware. The present
results suggest that substantial gains may be obtained by reducing redundancy at
the encoding stage before attempting larger or deeper quantum circuits.
This interpretation is consistent with broader NISQ-era guidance that useful
quantum performance depends critically on matching problem formulations to the
practical limits of existing devices \cite{preskill2018quantum,yarkoni2022quantum}.

The key question is how this approach scales with system size. The locality of
the rules is essential here. Because each rule acts on fixed-size graph
neighborhoods, the reduction can be applied to larger networks without first
enumerating all source--terminal paths. In hybrid workflows, this preprocessing
can reduce the size of the subgraphs subsequently mapped to quantum hardware.
Many practical workflows split the global problem into smaller
subproblems and solve them within a hybrid classical-quantum workflow. From
that perspective, improving the reliability of smaller routing graphs is useful
not only for the benchmark, but also as a building block for tackling larger
instances. Better small-graph solvers can therefore translate into better
performance in decomposition-based approaches to larger networks.

\section{Data and Code Availability}
The data and code supporting the findings of this study are available from the
corresponding author upon reasonable request.

\section{Competing Interests}
The authors are affiliated with QuantyMize Quantum Advance Ltd. and declare no
other competing interests. EGDT is a professor at Bar-Ilan University, Israel.

%Several extensions follow naturally from this work. First, the %reduced encoding
%should be tested on larger routing instances and on topologies with richer
%failure-correlation models. Second, the rule-based reduction can be compared not
%only with the benchmark quantum encoding but also with strong classical
%baselines, in order to clarify where quantum execution becomes competitive in
%practice. Third, future work should study the interaction between encoding
%reduction, error mitigation, and variational-circuit design, since these factors
%are likely to be complementary rather than independent.

%\section{References}
\begin{acknowledgments}
\textcopyright~2026 QuantyMize Quantum Advance Ltd.
The encoding methods described in this manuscript are the subject of a
pending patent application. For licensing and commercial-use inquiries,
please contact QuantyMize Quantum Advance Ltd. at \href{mailto:info@quantymize.com}{info@quantymize.com}.
\end{acknowledgments}

\end{document}